# 1,1-Diphenyl-2-picrylhydrazyl and superoxide anion radical scavenging activities of heterocyclic 2-oxo-1,2,3,4-tetrahydropyrimidines

Shahida Perveen[1,2], *Qurat-ul-Ain[3], Sarosh Iqbal[1], Sheeba Wajid[1], Khalid Muhammad khan[1], and Muhammad Iqbal Choudhary[1,3,4]

[1]H. E. J. Research Institute of Chemistry, International Center for Chemical and Biological Sciences, University of Karachi, Karachi-75270, Pakistan
[2]Department of Chemistry, Government Sadiq College Women University, Girls College Road-63100, Bahawalpur, Pakistan
[3]Dr. Panjwani Center for Molecular Medicine and Drug Research, International Center for Chemical and Biological Sciences, University of Karachi, Karachi-75270, Pakistan
[4]Department of Biochemistry, Faculty of Sciences, King Abdulaziz University, Jeddah-1412, Saudi Arabia

* Correspondence: quaratulain@iccs.edu

**Abstract:**

**Objective**

To investigate1,1-Diphenyl-2-picrylhydrazyl (DPPH) and superoxide radical (SOR) scavenging activities of 2-oxo-1,2,3,4-tetrahydropyrimidines derivatives. Free radicals are highly unstable and reactive molecules/atoms. In the body, free radicals form during normal and abnormal metabolism in the body and cause serious damage to other biomolecules through generating oxidative stress (OS). If free radicals induced OS does not neutralized properly, host multiple pathologies including several types of cancers.

**Method**

1,1-Diphenyl-2-picrylhydrazyl (DPPH) and superoxide (SOR) scavenging activities of 2-oxo-1,2,3,4-tetrahydropyrimidines derivatives were performed employing 1,1-Diphenyl-2-picrylhydrazyl (DPPH) and superoxide (SOR) scavenging assays in 96 well plate, ethanol was used as solvent to dissolve the compound.

**Results**

During current investigation, 2-oxo-1,2,3,4-tetrahydropyrimidine derivatives (**1-25**) were allowed to react with 1,1-Diphenyl-2-picrylhydrazyl (DPPH) and superoxide (SOR) radicals. Promising data were collected with $IC_{50}$ values ranging from $3.32 \pm 0.08$ μM to $167.31 \pm 0.74$ μM, as compared to positive reference compound quercetin ($IC_{50} = 94.1 \pm 1.2$ μM) in SOR assay. Whereas compound **13** exhibited significant activity in DPPH assay ($IC_{50} = 61.06 \pm 0.6$ μM), as compare to reference compound ascorbic acid ($IC_{50} = 40.1 \pm 1.1$ μM).

**Conclusions**

Hence, this preliminary study identifies a potent class of new antioxidant molecules that can serve as a lead towards oxidative stress related pathologies and cancers.

# Introduction

From ancient times, Free radicals are known to affect to human's life in various ways. Chemically, they are highly unstable and very reactive towards other molecules. In the body, not a single cell is prevented by such damage. Understanding the mechanisms of free radicals and related pathologies and their treatments attracted great attention by numerous researchers. Superoxide anion radical ($O_2^{·-}$) is most prevalent primary radical of human body. It is formed by one-electron transfer/reduction of molecular oxygen ($O_2$). The term "superoxide" coined for $O_2^{·-}$ as it is extraordinarily highly reactive, acts as a strong oxidizing agent. They continuously generate in cells under normal physiological conditions and lead to formation of broad range of other deadly free radicals collectively known as reactive oxygen species (ROS). It also known a powerful initiator of chain reactions in body). In the body, mainly two pathways generate superoxide anion radicals. First is the mitochondrial electron transport chain (ETC). During oxidative ATP production, some electrons "leak" to oxygen prematurely, forming superoxide oxygen, which has been reported in the range of serious health conditions (Valko, M et al 2007). The second source is NADPH oxidase enzyme that plays a crucial role in the immune system. Once activated, NADPH oxidase produce a burst of ROS that serve in killing invading pathogen. Un-encountered superoxide anion radicals lead to generation of other secondary radicals such as $OH^·$, $H_2O_2$, $RO^·$ and $ROO^·$. They all are oxygen-centered radicals. If these radicals are not handled properly, they accumulate in the body and attack various biomolecules in proximity such as proteins, lipids especially polyunsaturated fatty acids (PUFA) and nucleic acids DNA/RNA. oxidative damage to these molecules results in alteration their structure and function, hence, leads to stressful state in body "oxidative stress" (Mohana KN et al 2013). Also, this condition exaggerates other pathological conditions such as cancer, diabetes mellitus, rheumatoid arthritis, neurodegenerative and cardiac diseases, atherosclerosis, and causes early ageing (El-Bahr, 2013 and Basu Abhijit et al. 2022).

At the same time, our body has endowed with numerous endogenous antioxidant defense systems that may be enzymatic and non-enzymatic. This system provides prevention and protection against such unwanted accumulation oxidations and oxidative stress. More precisely, antioxidants work by neutralizing deadly free radicals and release oxidative stress.

Apart from endogenous antioxidants, exogenous antioxidants revolutionized the management of free radicals-induced bodily damages. They are known to neutralize radicals by accepting or donating electrons (Lobo, V. et al 2010). Some common exogenous antioxidants include vitamin E, vitamin C, Flavonoids, beta-carotene, and some omega FA etc. (Pham-Huy et al 2008). These strong antioxidants can be taken by diet (fruits/vegetables) or as supplements. Health and quality of life of an organism can be improved by essentially keeping the balance between oxidants and antioxidants.

Pyrimidines and fused pyrimidines represent a broad class of compounds, which have attracted great attention in medicinal chemistry. Extensive research is going on to discover new tetrahydropyrimidines due to their close structural features with clinically important dihydropyridine (calcium-channel blockers) (Fadda, AA et al 2013). Nitrogen containing Pyrimidine derivatives form a component in a number of useful drugs and are associated with many biological and therapeutical activities (Chaudhary A, et al 2011; Mohana KN et al 2013). Also they are reported against several biological activities such as anticancer (Steven, et al. 2010 and Perveen, S., etal, 2018), anti-inflammatory, Analgesic, ulcerogenic activity (El-Gazzar *et al*., 2007), anti-HIV (Gardelli, C et al 2007) etc.

Antioxidants are molecules that are proven to be important in treating multiple serious pathologies (cancer, diabetes mellitus etc.) and improving the quality of a patient's life. Recently our research group reported pyrimidine derivatives against xanthine oxidase (Zafar H et al 2018). Xanthine oxidase is an oxidative enzyme that generates $O_2^{-\bullet}$. Excellent results were observed with xanthine oxidase inhibition. In the present study, we evaluated pyrimidone derivatives **1-25** against DPPH and superoxide anion radicals *in vitro*. Derivatives **1-25** exhibited exciting results. Hence, this class of new compounds can serve as strong antioxidants and anti-inflammatory agents.

## Material and methods

NBT (Bio Basic Inc.), β-Nicotinamide adenine dinucleotide hydrogen (NADH) (Research Organics), DPPH (Sigma Chemicals), Phenazine methosulphate (Sigma Chemicals), Phosphate buffer solution, Ethanol and Dimethyl sulfoxide (Fisher Scientific), quercetin dihydrate (Bio-world). All chemicals used were of highest grade.

## Experimental

*Chemistry*

A series twenty-five 2-oxo-1, 2, 3, 4-tetrahydropyrimidines (**1-25**) was synthesized by combining urea, ethyl acetoacetate, and various aldehydes. Copper nitrate trihydrate was used as catalyst. Synthesis of 2-oxo-1, 2, 3, 4-tetrahydropyrimidines (**1-25**) was already reported by our research group (Iqbal S *et al* 2018). General procedure for synthesis of compounds (**1-25**) is presented below (scheme **1**).

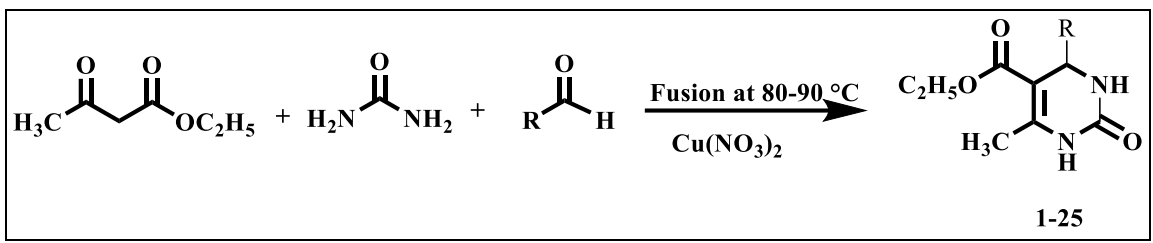

Scheme **1**: General procedure for synthesis of compounds **1-25**.

## Determination of DPPH radical scavenging activity

DPPH is a nitrogen-centered stable free radical. We performed DPPH assay in 96 well plate (Barakat, A.,2016). 100 µL of reaction mixture contains 5µL test sample with concentration 500 µM and 95 µL ethanolic solution of DPPH (300 µM). The plate was incubated at 37 °C for thirty minutes in dark. After incubation time, the absorbance was recorded at λmax=517 nm. The decrease in DPPH color (purple to yellowish) corresponds to reduction of DPPH radical with active compound. Vitamin C was used as reference compound (Ayaz A. et al 2014). The percent radical scavenging activity (% RSA) was calculated by the following equation (Eq **1**). DPPH radical scavenging activity was calculated using the following formula.
% DPPH radical scavenging activity of test sample = (1- Abs. of test sample/ Abs. of control) x 100

## Determination of Superoxide anion scavenging activity

Superoxide radicals were generated non-enzymatically by PMS/NADH/NBT system. Upon reduction of NBT, the dark blue colored formazan was measured according to the method (El Bakkali, M et al 2011). In above reaction β-NADH serves as electron donor.
Briefly, 200 µL of reaction mixture contained 10 µL of test sample (0.5 mM), 90 µL of sodium phosphate buffer (100 mM, pH 7.4); 40 µL of β-NADH (0.2 mM); and 40 µL of NBT (0.081 mM). Then 20 µL of PMS (0.008 mM) was added to start the reaction. The reaction was carried out in microtiter 96-well plate. Plate was incubated at 37 ˚C for 5 minutes.

The neutralization of superoxide radicals with active compounds was recorded by taking absorbance at 560 nm (molecular devices spectramax-384 spectrophotometer), against the DMSO treated negative control. Superoxide radical inhibitory activity was measured as inhibition of NBT's reduction in the presence of active compound. The decrease in absorbance corresponds to compound's activity. Quercetin dihydrate was used as positive control. The solutions of NADH, NBT and PMS were prepared in phosphate buffer. The test and standard samples were dissolved in DMSO (Hazra, B. et al 2008). Superoxide radical scavenging activity was calculated using the following formula.

% Superoxide radical scavenging activity of test sample = (1- Abs. of test sample/ Abs. of control) x 100

*Statistical analysis*

All data were expressed as mean values (n=3) with SEM (standard error of mean). Inhibitory concentrations by 50 % ($IC_{50}$) of each sample was calculated by EZ Fit enzyme kinetics software (Perrella Scientific, Inc. Amherst, U.S.A.).

## 4. Results

Twenty-five derivatives of heterocyclic 2-oxo-1,2,3,4-tetrahydropyrimidines were tested for antioxidant activity by employing two most routinely used SOR and DPPH radical bioassays. Different patterns of activities were observed due to presence of various substitutions as R group (Table-**1**). Interestingly, most of the currently investigated derivatives were highly active except compounds **5**, **7**, **12**, **20**, and **21** against SOR assay, however all derivatives remained inactive against DPPH radical except compound **13**. This observation depicts the involvement of different mechanisms attained to cope with different radicals. Pyrimidine derivatives **1-25** are substituted with various R groups, most small such as R=H (compound **1**) and bulky as R= anthracene moiety (compound **16**). Depending on such substituents attached to basic nucleus, varying degree of anti-radical activities were observed between $IC_{50}$ = 3.32 μM to 167.31 μM, in SOR assay as compared to reference compound, quercetin dihydrate ($IC_{50}$ = 94.1 ± 1.2 μM).

Compound **1**, the basic scaffold where R=H, showed potent superoxide anion radical scavenging activity with $IC_{50}$ = 24.81 ± 0.6 μM, which is far potent than positive control ($IC_{50}$ = 94.1 ± 1.2 μM). This observation indicates that the basic nucleus of this class contains this potency due to active groups present in heterocyclic ring, especially –NH groups, which most possibly take part in neutralization of $O_2\bullet-$. The addition of various R groups most probably affects the activity of –NH groups and thus the activity of each derivative. Compound **2** (R=$C_2H_5$) was highly active among the currently investigated series with $IC_{50}$ = 3.32 ± 0.08 μM. This observation suggests that addition of electron donating ethyl group might be responsible for increasing the potency of compound **2**. On the other hand, Compounds **3** (R= unsubstituted benzene ring) exhibited potent activity ($IC_{50}$ = 45.24±0.76 μM) as compared to reference compound ($IC_{50}$ = 94.1±1.2 μM). Compound **4**-**7** contained chloro group attached to benzene ring. The number and position of –Cl group substitutions makes drastic change in activities. –Cl group present at *para* position (compound **4**) exhibited remarkable anti-radical activity ($IC_{50}$ = 15.65±1.7 µM), as compared to compound **6** where the addition of another –Cl group at *ortho* position reduces the activity up to several folds ($IC_{50}$ = 167.31±0.7 µM). Whereas compounds **5** and **7** remain inactive. This observation suggested that para –Cl group seems enhance anti-radical activity on the other hand *ortho* Cl group is not effective rather declining the activity. Compound **9** was found to have slightly higher activity ($IC_{50}$ = 32.69±1.2 µM)) than compound **8** ($IC_{50}$ = 27.51±0.41 µM) therefore it can be suggested that *meta*-$NO_2$ group is most suitable than *para*-$NO_2$. Compounds **10-12** have –OH group at *meta*, *para*, and *ortho* positions, respectively. Interestingly, *o*–OH position was found not suitable for anti-radical activity. The order of activity is *para>meta>ortho* (Table-1). Compound **13** where *m*, p di-OH, group is present, showed significant superoxide anion radical scavenging activity ($IC_{50}$ = 68.91±6.3 µM), but lesser than mono-OH group. Compound **13** also exhibited promising activity against DPPH radicals with $IC_{50}$ = 61.06±0.6 µM as compared to vitamin C ($IC_{50}$ = 40.1 ± 1.1 µM), might be due to presence of two –OH on vicinal carbons.

Isopropyl substitution at para position of phenyl ring (compound **14**) showed significant activity ($IC_{50}$ = 55.03±0.76 µM), however this activity is several folds lesser than compound **15** (R= *p*-dimethylamine phenyl) which exhibited second most potent activity among the series ($IC_{50}$ = 13.0±0.8 µM).

R= 9-Anthracene in compound **16** was found to possess higher antioxidant activity in comparison with compounds **17** (R=2-naphthalene), and **18** (R=1-naphthalene) with IC50's values of 15.37±1.0 µM, 47.51±1.2 µM, 49.48±1.4 µM, respectively. This observation indicates the polynuclear aromatic substitutions especially anthracene, found to be effective in promoting antioxidant activities of under investigated pyrimidines. Compound **19** where R=3-Furan ring showed similar activity to compounds **17** and **18**.

Such potency of compounds, as we discussed earlier, might be due to resonance stabilization of resultant free radical by different electron donating substituents attached as R group. Whereas compounds **20** and **21** were inactive, showed percent RSA less than 50. Compound **22** (R = *o*, *m*, *p*-tri-$OCH_3$-phenyl) showed lesser activity ($IC_{50}$ = 77.7±2.56 µM), than compounds **23** (R=*o*-mono-$OCH_3$-phenyl) and **24** (R=*p*-mono-$OCH_3$-phenyl). The decline in antioxidant activity might be because of steric hindrance caused by three –$OCH_3$ groups on vicinal carbons. A slight increase in activity was observed with compound **25** where thio-methyl group is present at *para* position. In general, it can be said conclusively that potent activity of each active compound might attributed to their ability to form resonance-stabilized free radical (Khan, KM et al. 2012). Moreover, we also observed that electron-donating substituents are more effective in enhancing anti-radical activity of this class. Preliminary bio-evaluation of this class discovered **21** derivatives as potent SOR scavengers and only one found to be active in DPPH assay. Some more studies are still needed to establish their oxidative stress potential *in vivo*.

4. Discussions

In the present study, twenty-five heterocyclic 2-oxo-1,2,3,4-tetrahydropyrimidines (**1-25**) were evaluated for their antioxidant activities by superoxide anion and DPPH radical scavenging assays *in vitro*. Out of twenty-five investigated derivatives, twenty-one derivatives showed promising potential against superoxide anions varying from as potent as $IC_{50} = 3.32 \pm 0.08$ μM to moderate activity $IC_{50} = 167.31 \pm 0.7$ μM, depending on type

**Table 1:** Superoxide anion radical scavenging activities of 2-oxo-1,2,3,4-tetrahydropyrimidine derivatives.

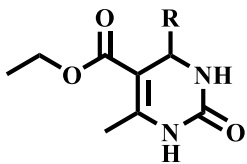

| S.# | R Group | RSA | $IC_{50}$ (μM) |
|---|---|---|---|
| 1. | H | 95% | 24.81±0.6 |
| 2. | $C_2H_5$ | 99% | 3.32±0.08 |
| 3. | phenyl | 96% | 45.24±0.76 |
| 4. | chlorophenyl | 97% | 15.65±1.7 |
| 5. | 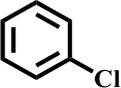 | 48% | N.D |
| 6. | 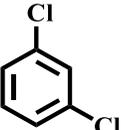 | 96% | 167.31±0.7 |
| 7. | 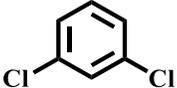 | NA | N.D |
| 8. | 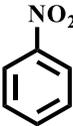 | 99% | 32.69±1.2 |
| 9. | 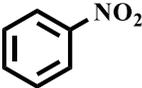 | 99% | 27.51±0.41 |
| 10. | 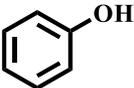 | 99% | 25.28±0.78 |
| 11. | 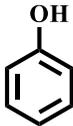 | 98% | 21.23±0.66 |
| 12. | 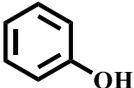 | N. A | N.D |
| 13. | 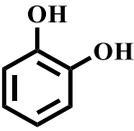 | 92% | 68.91±6.3 |
| 14. | 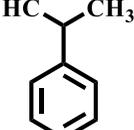 | 96% | 55.03±0.76 |
| 15. | 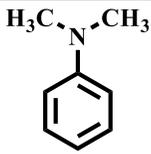 | 96% | 13.0±0.8 |
| 16. | 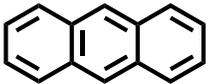 | 99% | 15.37±1.0 |
| 17. | 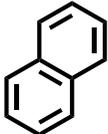 | 99% | 47.51±1.2 |
| 18. | 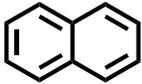 | 99% | 49.48±1.4 |
| 19. | 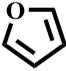 | 96 | 47.33±6.43 |
| 20. | 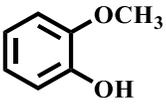 | NA | N.D |
| 21. | 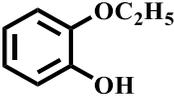 | NA | N.D |
| 22. | 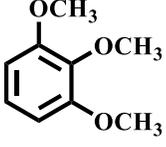 | 89% | 77.7±2.56 |
| 23. | 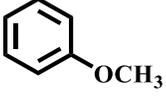 | 96% | 42.85±2.76 |
| 24. | 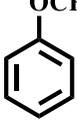 | 98% | 44.70±0.61 |

| 25. | 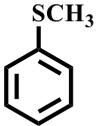 | 96% | 41.17±3.5 |
|---|---|---|---|
| **Std.** | Quercetin dihydrate | 96% | 94.1±1.2 |

$IC_{50}$ = mean of n=3±SEM; N.A= Not active (%RSA < 50); N.D= Not determined; std: standard

of substitutions. All compounds **1-25** were inactive in DPPH assay except compound **13**, exhibited strong DPPH activity. This implies that derivative **1-25** encounter different radicals with different mechanisms. Conclusively, this preliminary study identifies a new efficient anti-radical compound. These derivatives have potential to fight against deadly free radicals. However, more studies needed to find their effect against oxidative stress in cellular and animal models.

## Acknowledgement

We are thankful to Higher Education Commission Pakistan "Indigenous 5000 Fellowship Program" for carrying out this work.

## Conflict of Interest

Authors declare no conflict of interest.